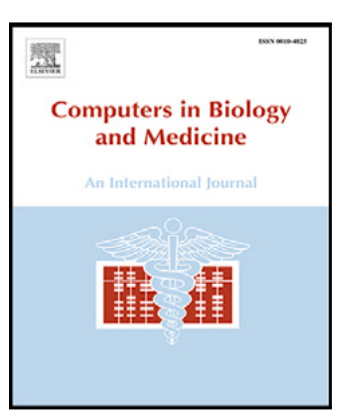

# UltraBones100k: A reliable automated labeling method and large-scale dataset for ultrasound-based bone surface extraction

Luohong Wu [a,*], Nicola A. Cavalcanti [a], Matthias Seibold [a], Giuseppe Loggia [b], Lisa Reissner [b], Jonas Hein [a,c], Silvan Beeler [b], Arnd Viehöfer [b], Stephan Wirth [b], Lilian Calvet [a], Philipp Fürnstahl [a]

[a] Research in Orthopedic Computer Science, Balgrist University Hospital, University of Zurich, Lengghalde 5, 8008, Zurich, Switzerland
[b] Department of Orthopaedics, Balgrist University Hospital, University of Zurich, Forchstrasse 340, 8008, Zurich, Switzerland
[c] Computer Vision and Geometry Group, ETH Zurich, Ramistrasse 101, 8092, Zurich, Switzerland

## ARTICLE INFO

## ABSTRACT

**Background:** Ultrasound-based bone surface segmentation is crucial in computer-assisted orthopedic surgery. However, ultrasound images have limitations, including a low signal-to-noise ratio, acoustic shadowing, and speckle noise, which make interpretation difficult. Existing deep learning models for bone segmentation rely primarily on costly manual labeling by experts, limiting dataset size and model generalizability. Additionally, the complexity of ultrasound physics and acoustic shadow makes the images difficult for humans to interpret, leading to incomplete labels in low-intensity and anechoic regions and limiting model performance. To advance the state-of-the-art in ultrasound bone segmentation and establish effective model benchmarks, larger and higher-quality datasets are needed.

**Methods:** We propose a methodology for collecting ex-vivo ultrasound datasets with automatically generated bone labels, including anechoic regions. The proposed labels are derived by accurately superimposing tracked bone Computed Tomography (CT) models onto the tracked ultrasound images. These initial labels are refined to account for ultrasound physics. To clinically evaluate the proposed method, an expert physician from our university hospital specialized in orthopedic sonography assessed the quality of the generated bone labels. A neural network for bone segmentation is trained on the collected dataset and its predictions are compared to expert manual labels, evaluating accuracy, completeness, and F1-score.

**Results:** We collected UltraBones100k, the largest known dataset comprising 100k ex-vivo ultrasound images of human lower limbs with bone annotations, specifically targeting the fibula, tibia, and foot bones. A Wilcoxon signed-rank test with Bonferroni correction confirmed that the bone alignment after our optimization pipeline significantly improved the quality of bone labeling ($p < 0.001$). The model trained on UltraBones100k consistently outperforms manual labeling in all metrics, particularly in low-intensity regions (at a distance threshold of 0.5 mm: 320% improvement in completeness, 27.4% improvement in accuracy, and 197% improvement in F1 score)

**Conclusion:** This work is promising to facilitate research and clinical translation of ultrasound imaging in computer-assisted interventions, particularly for applications such as 2D bone segmentation, 3D bone surface reconstruction, and multi-modality bone registration.

## 1. Introduction

Ultrasound is a well-established imaging technology widely used in the clinical diagnosis of musculoskeletal pathologies, including bone and soft tissue injuries [1]. It also facilitates dynamic assessments of joint and tendon movements, stability [2], and the detection of fracture healing, structural abnormalities, infections, ligament injuries, nerve compression, and mechanical impingement [3]. As evidenced by a substantial body of literature [4–8], the automated segmentation and 3D reconstruction of bone surfaces from ultrasound images are gaining increasing significance as a powerful tool for interventional Computer-Assisted Orthopedic Surgery (CAOS) systems. Compared to ionizing imaging modalities like fluoroscopy, ultrasound offers advantages in improving guidance and navigation during interventional and surgical procedures [4], including trauma treatment [6], needle placement [5], and robotic spine surgery [9].

* Correspondence to: Lengghalde 5, 8008, Zurich, Switzerland
*E-mail address:* luohong.wu@balgrist.ch (L. Wu).

Accurate 3D modeling of bone geometry requires precise 2D identification and segmentation of bone anatomies in ultrasound images, which relies heavily on high-quality labels produced by expert professionals [10–13]. Due to the highest acoustic impedance compared to surrounding soft tissues, bones typically appear as high-intensity regions when the ultrasound probe is perpendicular to the bone surface. However, when the probe is positioned at a larger oblique angle to the bone surface, bones can appear as low-intensity regions as most portion of acoustic energy is reflected into surrounding soft tissues instead of back to the probe. These low-intensity regions can only be inferred by surgeons through anatomical knowledge but cannot be accurately labeled. The same applies to fully anechoic regions formed by the complete absorption of the ultrasound signal in the overlying bone layers.

To address these challenges, various medical imaging methods based on handcrafted features have been developed to enhance the interpretation of ultrasound images [14,15]. Recently, deep learning-based bone segmentation methods have shown promising improvements in robustness, accuracy, and processing speed [7,8,16–20]. However, the performance of such models relies on the quality of the manually labeled training dataset. Manual bone labeling is time-consuming, and requires specialized knowledge of ultrasound imaging and bone anatomy, which cannot be produced on a large scale if relying solely on human experts. The resulting datasets are often small, ranging from dozens to a few thousands labeled slices. This limits the performance and generalizability of the trained models. Additionally, most existing methods train and validate their models on non-public datasets and use non-standardized evaluation metrics, making it difficult to compare and benchmark the performance of different models across studies [21,22]. The absence of large-scale, accurately labeled public datasets has been a critical obstacle for the community in developing more advanced models.

Bone surface registration between CT and ultrasound has been employed to evaluate the accuracy of bone surface extraction in ultrasound images [6]. However, achieving accurate and fully automatic bone labeling based on CT–ultrasound alignment remains challenging due to the high precision required. In this work, we propose a methodology that enables the collection of large-scale ultrasound datasets with automatically generated bone labels. Instead of relying on manual labeling, our approach utilizes optical tracking to align a 3D bone model, extracted from a preoperative CT scan, with tracked ultrasound images. The initial alignment is further refined using our proposed intensity-based optimization strategy for each frame, rather than relying on traditional global optimization [23]. This approach enhances the precision and completeness of bone labels in ultrasound images. With this methodology, we collected a dataset of 100k annotated ultrasound images with bone labels from 14 human cadaveric lower legs. A deep learning model for bone segmentation is trained on the resulting UltraBones100k dataset, which significantly outperforms manual segmentation performed by an expert surgeon.

To ensure reproducibility of the results, the dataset and pretrained model weights, have been made publicly available[1].

## 2. Related work

**Ultrasound Bone Segmentation.** Numerous intensity-based bone segmentation methods have been developed [21,22]. However, intensity-based methods face challenges in terms of robustness and accuracy. To mitigate these issues, frequency-based methods have been introduced, including phase symmetry and structured phase symmetry techniques [14,15,24], which aim to enhance bone boundaries. With the advent of deep learning, neural network models such as CNNs, U-nets, and GANs have been employed for bone segmentation in ultrasound images, targeting various anatomical regions such as the spine [7,16,17], pelvis [8,18], femur [8,19,20], tibia [8,19,20], fibula [19], and radius [19,20]. These deep learning methods have demonstrated superior performance over traditional approaches in terms of accuracy, robustness, and runtime. For example, average bone surface distance errors have been reported to range from 0.2 mm to 1.7 mm [7,18,20,25]. Based on accurate 2D bone segmentation results, the reconstructed bone surfaces achieve sub-millimetric one-sided Chamfer Distance (CD) [11]. However, these studies rely on manually annotated bone labels from expert surgeons. This process is time-consuming, costly, and requires specialized knowledge of ultrasound imaging and bone anatomy. In addition, different surgeons may also provide different labels for the same ultrasound image, leading to inconsistencies. As a result, the quality and the size of training datasets is limited, typically ranging from dozens to a few thousands of labeled images [22]. These constraints reduce model performance and generalizability. In addition, the absence of standardized evaluation metrics, with over 18 different metrics in use [21], further hinders effective model benchmarking and comparison.

**Automatic labeling.** Optical tracking markers have been applied in the medical imaging community for collecting datasets with automatically generated labels, such as tracked spine CT models [26] and surgical instruments [27]. So far, one previous work has attempted to automatically generate bone labels in ultrasound images using a tracked bone CT model [12]. However, their approach led to significant alignment errors due to error accumulation, ultimately requiring a fallback to manual labeling. Traditional image-based multi-modal registration methods, such as CT-US registration in Imfusion Suite,[2] compute a transformation matrix to globally register one modality to another [23]. These methods are highly dependent on initialization and global intensity distribution, which can lead to local minima. As we will demonstrate, these methods do not provide sufficient accuracy for automatic labeling. In contrast, our proposed method optimizes the alignment for each frame locally, resulting in more accurate bone labeling.

## 3. Method

### 3.1. Dataset collection and automatic bone labeling

**Cadaveric Specimens.** The donated specimens used in this study were fresh-frozen human cadaveric lower limbs, obtained from Science Care (Phoenix, Arizona, USA). The donors cover different physical characteristics, with heights ranging from 155 to 185 cm, weights from 59 to 99 kg, and BMI values between 19 and 30. The age range spans from 41 to 76 years, and the dataset includes 7 male and 7 female specimens. Furthermore, arthritis status is documented, with some specimens having a history of falls or related conditions. Following donor death, the limbs were frozen shortly thereafter and subsequently stored in a dedicated laboratory freezer at −22 °C. Prior to use in the experimental setup, the specimens were thawed at room temperature (∼20 °C) for 36 h, in accordance with an internal standardized wetlab protocol. This thawing process was performed to ensure complete tissue thawing while minimizing tissue degradation and avoiding thermal artifacts that could affect ultrasound imaging. To maintain tissue integrity, all specimens were handled exclusively by wetlab-trained specialist medical professionals. Thus, all ultrasound scanning was conducted on fully thawed, all ultrasound scanning was performed on fully thawed, fresh-frozen human tissue under controlled and standardized conditions, ensuring high fidelity and reproducibility of imaging results.

[1] https://github.com/luohwu/UltraBones100k.git.

[2] https://www.imfusion.com/.

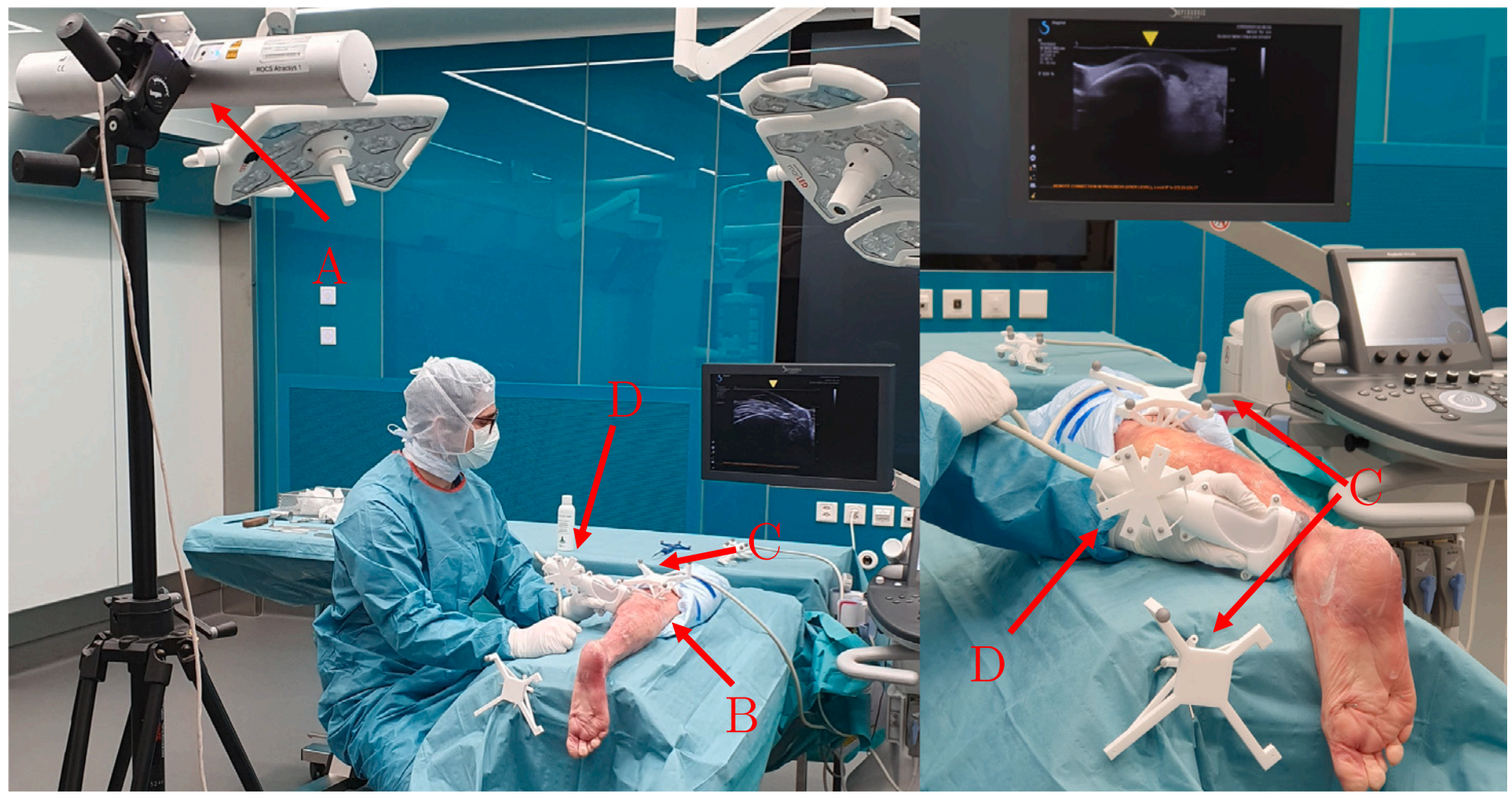

**Fig. 1.** The dataset acquisition setup includes an optical tracking system (A), a human cadaveric specimen (B) with mounted optical markers (C), and an tracked and calibrated ultrasound device (D).

**Acquisition Setup**. Our dataset acquisition setup using human cadaveric specimens is illustrated in Fig. 1. Ethics approval has been granted for our study by the Zurich cantonal ethics committee under BASEC Nr. 2023-01681. Prior to data acquisition, K-wires with a diameter of 2.5 mm and a length of 150 mm (DePuy Synthes, USA), along with custom-made 3D-printed marker mounts equipped with infrared-sensitive fiducial markers, were mounted to each specimen. The K-wires were used to stabilize the bones of interest, including the lower leg and foot, to prevent bone movement during scanning. Optical markers were tracked by an optical tracking system (FusionTrack 500, Atracsys, Switzerland). A CT scan was performed on each specimen with an image dimension of 512 × 512 pixels, an in-plane pixel spacing of 0.839 mm × 0.839 mm, and a slice thickness of 0.6 mm (NAEOTOM Alpha, Siemens, Germany). CT Bone segmentation was generated using the region growing and thresholding functionalities in Mimics (Materialise, Leuven, Belgium). For data acquisition, each specimen was scanned nine times using an ultrasound device, the Aixplorer Ultimate (SuperSonic Imagine, Aix-en-Provence, France), equipped with an SL18-5 linear probe (SuperSonic Imagine, Aix-en-Provence, France). The probe features 256 elements, an acoustic footprint of 50 mm, and a bandwidth range of 5–18 MHz. Details of the imaging configuration are presented in Table 1. Multiple 3D-printed optical marker arrays are attached on the probe, optimizing the tracking performance in 6 degrees-of-freedom in the target space. In addition, to ensure proper contact between the probe and the specimens for high-quality ultrasound imaging, a gel pad (aiSon$^{TM}$ Technologies, Zurich, Switzerland)[3] was attached to the top of the ultrasound probe. The scanning procedure consisted of three forward and backward passes along the longitudinal axis of the tibia, three along the fibula, and three covering the foot (including the bones of the ankle joint such as the calcaneus and talus), aiming to maximize bone surface coverage. During the scanning, the movement of the ultrasound probe is not limited explicitly. The clinician can keep scanning until our system provide a signal indicating that the probe is outside of the workspace visible by the tracking system, which is considered as the limitation along that direction. Ultrasound images, along with the tracking data of the ultrasound probe and the specimen, were recorded with corresponding timestamps.

**System Calibration**. To determine the coordinates of the tracking fiducials on the specimen marker within the CT space, virtual spheres are created to match the size of the physical fiducials in MeshLab [28]. These virtual spheres are initially manually aligned with the corresponding fiducials in the segmented CT model, followed by refinement using the point-to-point Iterative Closest Point algorithm [29]. The resulting registration transformation matrix allows for the accurate calculation of relative poses between the ultrasound probe and the bone CT model. However, converting each pixel of an ultrasound image into its corresponding point in the CT space requires accurate spatial calibration and accurate tracking of the ultrasound probe. In addition, temporal calibration is necessary to synchronize the ultrasound images with the tracking data. Following the convention in existing works [30,31], we define the origin of the ultrasound image at the top-left corner. The transformation chain across all components is illustrated in Fig. 2. The pixels of the ultrasound image at timestamp $t$ are defined in homogeneous coordinates as $\mathcal{P}^t = \left\{ [u, v, 0, 1]^T \mid u \in [0, W), v \in [0, H) \right\}$, where $W$ and $H$ represent the image width and height, respectively. The corresponding point cloud in CT space, $\mathcal{U}^t = \left\{ [x_j, y_j, z_j, 1]^T \mid j \in [0, H \times W) \right\}$, is computed as

$$\mathcal{U}^t = \mathrm{T}^{t+\Delta t}_{\mathrm{CT}\leftarrow\mathrm{OT}} \cdot \mathrm{T}^{t+\Delta t}_{\mathrm{OT}\leftarrow\mathrm{US}} \cdot \mathrm{T}_{\mathrm{US}\leftarrow\mathrm{IP}} \cdot \mathrm{T}_{\mathrm{S}} \cdot \mathcal{P}^t = \mathrm{T}^{t+\Delta t}_{\mathrm{CT}\leftarrow\mathrm{US}} \cdot \mathrm{T}_{\mathrm{US}\leftarrow\mathrm{IP}} \cdot \mathrm{T}_{\mathrm{S}} \cdot \mathcal{P}^t \tag{1}$$

with:

- $\mathrm{T}_{\mathrm{S}} = \begin{bmatrix} s_x & 0 & 0 & 0 \\ 0 & s_y & 0 & 0 \\ 0 & 0 & 1 & 0 \\ 0 & 0 & 0 & 1 \end{bmatrix}$, where $s_x$ is the width of one pixel in mm, and $s_y$ is the height of one pixel in mm. $s_x$ and $s_y$ are known from the ultrasound device.
- $\mathrm{T}_{\mathrm{US}\leftarrow\mathrm{IP}} = \begin{bmatrix} r_{11} & r_{12} & r_{13} & d_x \\ r_{21} & r_{22} & r_{23} & d_y \\ r_{31} & r_{32} & r_{33} & d_z \\ 0 & 0 & 0 & 1 \end{bmatrix}$, belonging to the 3D Special Euclidean group SE(3), is the transformation matrix that maps 3D points from the coordinate system of the image plane (origin: top-left corner) to the coordinate system of the marker attached to the ultrasound probe. The determination of this matrix refers to the process of spatial ultrasound calibration [30,31]. Note that a transformation matrix $\mathrm{T}(\boldsymbol{\theta}, \mathbf{d})$ can be defined using the Euler rotation angles $\boldsymbol{\theta}$ and the translation vector $\mathbf{d}$, which together specify the relationship between two coordinate systems.
- $\mathrm{T}^{t+\Delta t}_{\mathrm{OT}\leftarrow\mathrm{US}} \in \mathrm{SE}(3)$ is the transformation matrix at timestamp $t + \Delta t$ that transforms 3D points from the coordinate system of the ultrasound probe marker to the coordinate system of the optical tracking system. As the probe moves during dataset collection, $\mathrm{T}_{O\leftarrow US}$ changes over time instead of a constant matrix. At each timestamp $t$, two types of data were collected: ultrasound image data ($\mathcal{P}^t$) and tracking data from multiple markers ($\mathrm{T}^t$). Due to differences in sensor characteristics, such as frame rates and data acquisition timings, $\mathcal{P}^t$ and $\mathrm{T}^t$ are not inherently synchronized, resulting in a temporal offset denoted as $\Delta t$. Determining $\Delta t$ is thus the process of temporal calibration. In our setup, the tracking data has a higher temporal resolution than the ultrasound data ($\Delta t < 0$), as forming a B-mode ultrasound image requires additional time to process the original acoustic signals.
- $\mathrm{T}^{t+\Delta t}_{\mathrm{CT}\leftarrow\mathrm{OT}} \in \mathrm{SE}(3)$ is the transformation matrix at timestamp $t + \Delta t$ from the coordinate system of the optical tracking system to the coordinate system of the CT scan, which can be determined using the marker attached to the cadaveric specimen.
- For simplicity, we denote $\mathrm{T}^{t+\Delta t}_{\mathrm{CT}\leftarrow\mathrm{US}}(\boldsymbol{\theta}, \mathbf{d}) = \mathrm{T}^{t+\Delta t}_{\mathrm{CT}\leftarrow\mathrm{OT}}(\boldsymbol{\theta}_1, \mathbf{d}_1) \cdot \mathrm{T}^{t+\Delta t}_{\mathrm{OT}\leftarrow\mathrm{US}}(\boldsymbol{\theta}_2, \mathbf{d}_2)$ as the direct transformation from the ultrasound marker to the CT space.

**Table 1**
Detailed imaging configuration. Initially, the device's built-in preset "Orthopedics" was used for setup. Subsequently, the clinician fine-tuned the parameters listed in the table.

| Gain | 48% | Tissue tuner | 1540 m/s |
|---|---|---|---|
| Max depth | 5 cm | Focus depth | 2.5 cm |
| Dynamic range | 62 dB | Sector size | Large |
| Harmonic imaging | Disabled | Super compound | Enabled |
| Needle PLUS | Disabled | Res/Gen/Pen | Res |

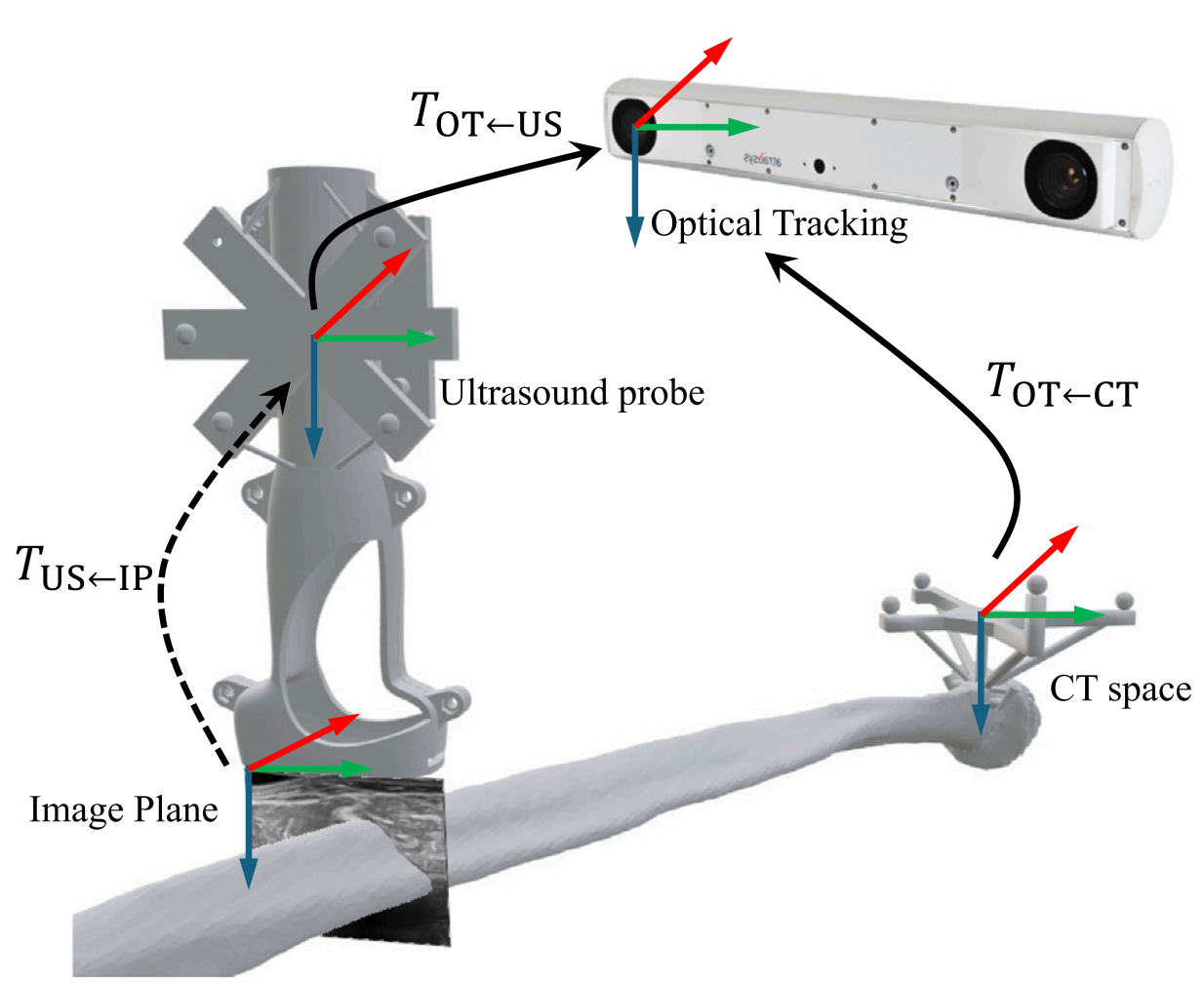

**Fig. 2.** Diagram explaining the transformation chain across all components. Solid black arrows represent known transformations, which are directly measured by the optical tracking system. The dashed-line arrow represents the unknown transformation from the image plane to the ultrasound probe, which will be determined through calibration.

[3] https://www.aisontechnologies.com/aison-focus-standoff-pad/.

Unlike previous studies that treated spatial and temporal calibration independently [30,31], we performed both calibrations simultaneously using a Z-phantom [31]. The temporal offset $\Delta t$ was sampled over the range $[-300, 0)$ ms with a step size of 0.1 ms. For each $\Delta t$, $\mathrm{T}_{\mathrm{US}\leftarrow\mathrm{IP}}$ was computed, and the corresponding 3D localization recorded. The optimal parameters were identified by minimizing the 3D localization error. Our method achieved a 3D localization error of 1.07 mm, outperforming previously reported errors ($\approx$1.5 mm) [30,31]. With precise calibration results, 2D pixel coordinates can be accurately mapped to 3D points in the CT space. For simplicity, we denote the pixel corresponding to the point $\mathbf{u}^t_i \in \mathcal{U}^t$ as $\mathbf{p}^t_i \in \mathcal{P}^t$, and define $f(\mathbf{u}^t_i)$ as the intensity value of $\mathbf{p}^t_i$ in the original ultrasound image.

**Initial Bone Labeling** Points are sampled uniformly over the bone CT model to generate the CT point cloud $\mathcal{C}$. Both $\mathcal{C}$ and $\mathcal{U}^t$ (with original intensity values) are rendered in Fig. 3. The CT-ultrasound bone surface intersection is defined as $\mathcal{I}^t = \mathcal{C} \cap \mathcal{U}^t = \left\{ \mathbf{u} \in \mathcal{U}^t \mid \min_{\mathbf{c}\in\mathcal{C}} \|\mathbf{u} - \mathbf{c}\| \leq \gamma \right\}$ where $\gamma$ is the parameter defining the thickness of the intersection ($\gamma = 0.3$ mm determined empirically). Ideally, $\mathcal{I}^t$ can be projected back onto the ultrasound image, forming a complete bone label. However, as also reported previously [12], mismatches (1 to 2 mm) often occur due to accumulated errors, including tracking error and calibration error (see Fig. 3). Although these mismatches are typically below 2 mm, they are noticeable given low pixel spacing of ultrasound images, which can be as small as 0.05 mm per pixel.

**Optimization Step for Refined Alignment.** Bone contours that are roughly perpendicular to the ultrasound probe typically appear as high-intensity regions in ultrasound images due to the underlying ultrasound physics. Based on this principle, we propose an intensity-based optimization method designed to refine the location of CT-derived labels. The proposed optimization relies on the following assumptions: (i) the main error sources are tracking and calibration errors, as the only undetermined variables in Eq. (1) are the tracking matrices ($\mathrm{T}^{t+\Delta t}_{\mathrm{CT}\leftarrow\mathrm{OT}}$ and $\mathrm{T}^{t+\Delta t}_{\mathrm{OT}\leftarrow\mathrm{US}}$) and the calibration matrix ($\mathrm{T}_{\mathrm{US}\leftarrow\mathrm{IP}}$); (ii) pixels with high intensity in the vicinity of the non-corrected labels mostly correspond to the true bone contour. Beneath the bone shadow, there are almost uniformly dark regions, as bones possess the highest acoustic impedance compared to the surrounding soft tissues, causing most acoustic waves to be reflected back; (iii) the CT-ultrasound alignment error does not exceed a mean distance of 2 mm, given that the optical tracking system offers sub-millimeter accuracy[4] and the calibration matrix achieves a 3D localization error of 1.07 mm. Furthermore, a random image sampled from each ultrasound sweep to measure mismatch yielded a mean error of $1.13 \pm 0.606$ mm.

The key principle of our optimization method is to add small perturbations ($\boldsymbol{\theta}_\epsilon$ and $\mathbf{d}_\epsilon$) to each ultrasound image pose in order to maximize the intensity of pixels at the intersection between the ultrasound image and bone CT model while minimizing deviations from the recorded tracking data. We define the perturbed point cloud as $\mathcal{U}^t(\boldsymbol{\theta}_\epsilon, \mathbf{d}_\epsilon) = \mathrm{T}^{t+\Delta t}_{\mathrm{CT}\leftarrow\mathrm{US}}(\boldsymbol{\theta} + \boldsymbol{\theta}_\epsilon, \mathbf{d} + \mathbf{d}_\epsilon) \cdot \mathrm{T}_{\mathrm{US}\leftarrow\mathrm{IP}} \cdot \mathrm{T}_{\mathrm{S}} \cdot \mathcal{P}^t$ and the perturbed intersection as $\mathcal{I}^t(\boldsymbol{\theta}_\epsilon, \mathbf{d}_\epsilon) = \mathcal{C} \cap \mathcal{U}^t(\boldsymbol{\theta}_\epsilon, \mathbf{d}_\epsilon)$. The proposed solution is written as:

$$\underset{\boldsymbol{\theta}_\epsilon, \mathbf{d}_\epsilon \in [-1,1]^3}{\arg\max} \frac{\sum_{\mathbf{u}\in\mathcal{I}^t(\boldsymbol{\theta}_\epsilon,\mathbf{d}_\epsilon)} f(\mathbf{u})}{|\mathcal{I}^t(\boldsymbol{\theta}_\epsilon,\mathbf{d}_\epsilon)|} - \lambda \frac{\sum_i \|\mathrm{T}(\boldsymbol{\theta},\mathbf{d})\cdot\mathbf{F}_i - \mathrm{T}(\boldsymbol{\theta}+\boldsymbol{\theta}_\epsilon,\mathbf{d}+\mathbf{d}_\epsilon)\cdot\mathbf{F}_i\|_2}{N} \tag{2}$$

$\mathbf{F}_i$ denotes the homogeneous coordinate vector of the $i$th tracking fiducial of the ultrasound probe marker, with $N$ fiducials in total. The first term of Eq. (2) represents the mean intensity values of points in $\mathcal{I}^t(\boldsymbol{\theta}_\epsilon, \mathbf{d}_\epsilon)$ under the corrected image pose. The second term represents the mean deviation of the tracking fiducials induced by the pose correction in terms of Euclidean distance. By leveraging the bounded set and regularization term, our approach ensures the solution remains close to the initial tracking data. The weighting parameter $\lambda$ is empirically determined from the range [0.01, 0.1, 1, 10, 100,1000]. The optimization problem is solved using the differential evolution method [32]. Each $\lambda$ value was tested on randomly sampled frames, and the results were evaluated quantitatively to select the optimal $\lambda$. To identify failed optimizations, a validation step is performed. Specifically, ultrasound images with a mean fiducial location correction greater than 1 mm after optimization are discarded, as the tracking device provides sub-millimetric accuracy. After the validation step, approximately 18.7% frames are discarded, with the mean fiducial location correction of the remaining frames being 0.481 mm. An example of a full bone label refined using the proposed optimization is shown in Fig. 3. To account the bone shadow effect, bone regions within the shadow (blue region in Fig. 3) are discarded. In addition, we project the normal vectors from the bone CT model onto the ultrasound images and compute the incidence angles $\alpha$ between the projected normal vectors and the vertical scan lines, as shown in Fig. 3. According to ultrasound imaging physics, the intensity of bone pixels decreases as $\alpha$ increases, with $\alpha > 90°$ being a critical boundary. In this study, we keep bone pixels with $\alpha \leq 85°$ to avoid boundary cases. The final bone label is shown in Fig. 3. More examples of bone labels before and after the optimization are shown in Fig. 4.

[4] https://atracsys.com/fusiontrack-500/.

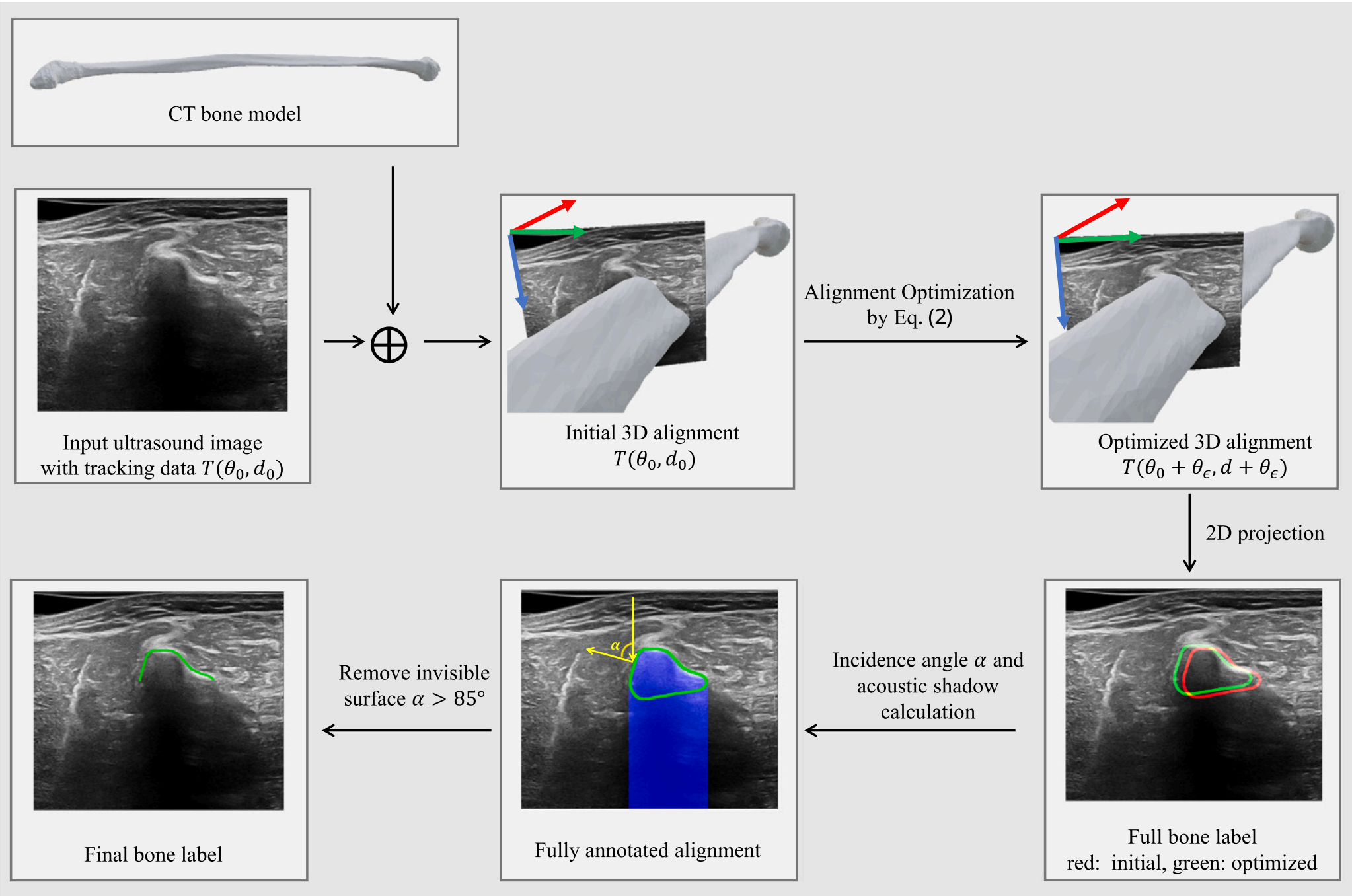

**Fig. 3.** Overview of the automated 2D bone labeling pipeline. By leveraging 3D spatial data, the ultrasound image and CT bone model are overlaid to produce an initial bone alignment (red). Optimization of Eq. (2) refines the alignment (green). Following the refinement, the bone shadow (blue) and incidence angles (yellow) are computed. Using ultrasound physics, invisible bone surface regions are excluded, resulting in the final bone label.

**Clinical Evaluation.** To clinically evaluate the quality of bone annotations following our optimization pipeline, tow expert physician specialized in orthopedic sonography assessed the quality of both the initial and optimized bone annotations. For comparison with State-of-the-Art (SOTA), we also employed the traditional "ultrasound to CT registration" algorithm available in the ImFusion Suite[5] (ImFusion, Munich, Germany) to compute a global transformation matrix for registering the ultrasound data to the CT volume (the default parameter setting yielded the best results in our experiments), rather than optimizing the alignment for each frame as our method does. In our experiment, one frame was randomly sampled from each ultrasound sweep, resulting in a total of 119 frames. Bone annotations were performed using three alignment methods: the initial alignment, the optimized alignment derived from our method, and the bone alignment obtained after the CT-US registration step. The resulting bone alignment outputs from these three methods were presented to an expert physician specialized in orthopedic sonography in random order. The expert evaluated the quality of each bone alignment using the following criteria:

- **0:** Poor alignment, with large sections showing visible errors exceeding 3 mm.
- **1:** Moderate alignment, characterized by some sections with visible errors between 1–3 mm.
- **2:** Good alignment, exhibiting only minor visible errors below 1 mm.
- **3:** Excellent alignment, with no visible errors.

A screenshot of the rating software interface is shown in Fig. 5. A scale indicating 1, 2, and 3 mm is displayed in the top-left corner of each image.

### 3.2. Bone segmentation

**Segmentation Mode.** In our study, 14 cadavers were scanned, resulting a dataset of over 100k ultrasound images with bone labels. We split the dataset into 10 training specimens, 1 validation specimen, and 3 test specimens (around 25k ultrasound images). The training and validation sets are further split using cross-validation. A U-net with ResNet encoders is trained on UltraBones100k to automatically segment bones in ultrasound images [33,34]. The loss function is a weighted combination of Dice Loss (weight=1) and Binary Cross Entropy Loss (weight=1). We trained the model using the Adam optimizer with $\beta = (0.9, 0.999)$, starting with an initial learning rate of 1e–5, which decayed by 0.9 every 10 epochs. The training process ran for 100 epochs on an NVIDIA V100 GPU, with a batch size of 32. The code and pretrained model have been made publicly available[6] .

**Manual annotation.** To compare the automatic bone labeling by the trained model with manual labeling, 700 ultrasound images were randomly sampled from the test set and manually labeled by an expert surgeon. The surgeon identified the hyperechoic bone surface layer and its corresponding acoustic shadow based on the following knowledge about ultrasound imaging and anatomical structures. Bone surfaces in ultrasound images are identified as a hyperechoic (bright) layer resulting from the significant reflection of ultrasound waves at the bone-tissue interface. This bright layer is typically accompanied by a characteristic acoustic shadow extending posteriorly, as the underlying bone attenuates or reflects the majority of the ultrasound signal. The anatomical region being scanned significantly influences the clarity and confidence of bone surface identification. Larger bones (e.g., >1 cm in diameter) with smooth, superficial surfaces are more readily identified due to their distinct echogenicity and reduced interference from surrounding tissue artifacts or acoustic noise. In contrast, smaller or deeper bones, or those with irregular surfaces, may present more ambiguous features. Experienced operators rely on a combination of real-time image feedback and their knowledge of anatomical landmarks to delineate tissue structures, including bone surfaces. This process underscores the operator-dependent nature of ultrasound-based diagnostics in clinical practice. For labeling purposes, the ImFusion Label software[7] (ImFusion, Munich, Germany) was employed to annotate the bone surface layer on a pixel-wise basis within the acquired images. This software allows precise manual delineation of the bone surface, facilitating the creation of ground-truth labels for subsequent analysis. **Evaluation Metrics.** In previous studies [21,22], over 18 different metrics were used for evaluating bone segmentation in ultrasound images, including intersection-based (e.g., Dice score) and distance-based metrics (e.g., one-sided CD), which complicates fair benchmarking of models. Bone labels in ultrasound are thin, even small shifts can drastically reduce intersection-based scores. While CD is easy to interpret, it struggles with false positives. To address this, we propose using distance-thresholded accuracy, completeness, and F1-score, which balance intersection-based and distance-based evaluations. Accuracy represents the fraction of correct predictions out of all predictions at a certain distance threshold, while completeness measures the fraction of correct predictions relative to the ground truth. Particularly, our evaluation is based on the CT-derived labels, allowing to compare the performance between automatic segmentation and manual labeling. Specifically, for a pixel in the predicted bone label $\hat{\mathbf{x}} \in \mathcal{P}$, its distance to the ground truth $\mathcal{G}$ is defined as follows:

$$d_{\hat{\mathbf{x}} \to \mathcal{G}} = \min_{\mathbf{x} \in \mathcal{G}} \|\hat{\mathbf{x}} - \mathbf{x}\|, \tag{3}$$

[5] https://www.imfusion.com/.

[6] https://github.com/luohwu/UltraBones100k.git.

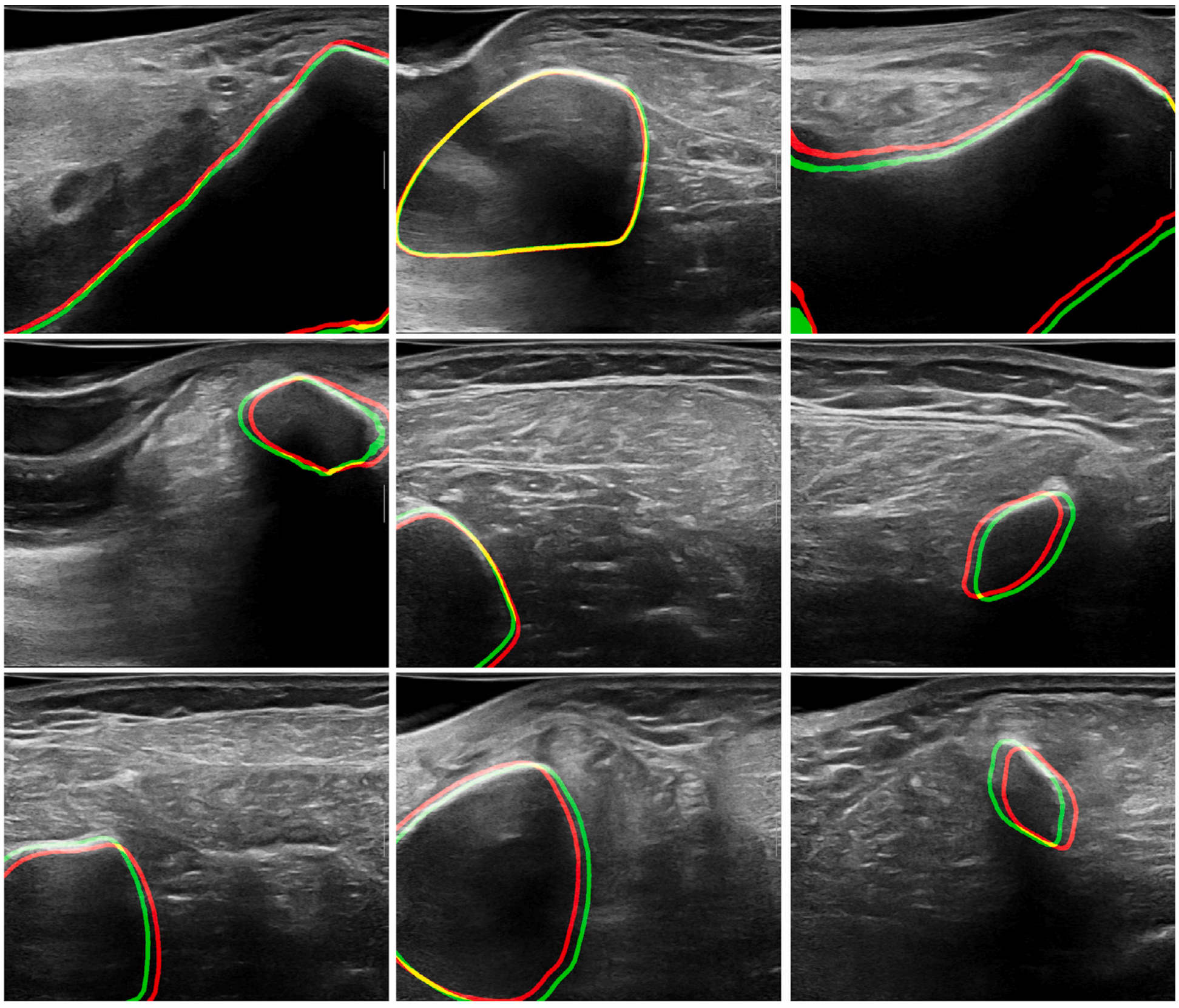

**Fig. 4.** Comparison results of bone annotation before and after optimization. Red regions show the initial alignment, green regions show the refined alignment, yellow regions are the overlap.

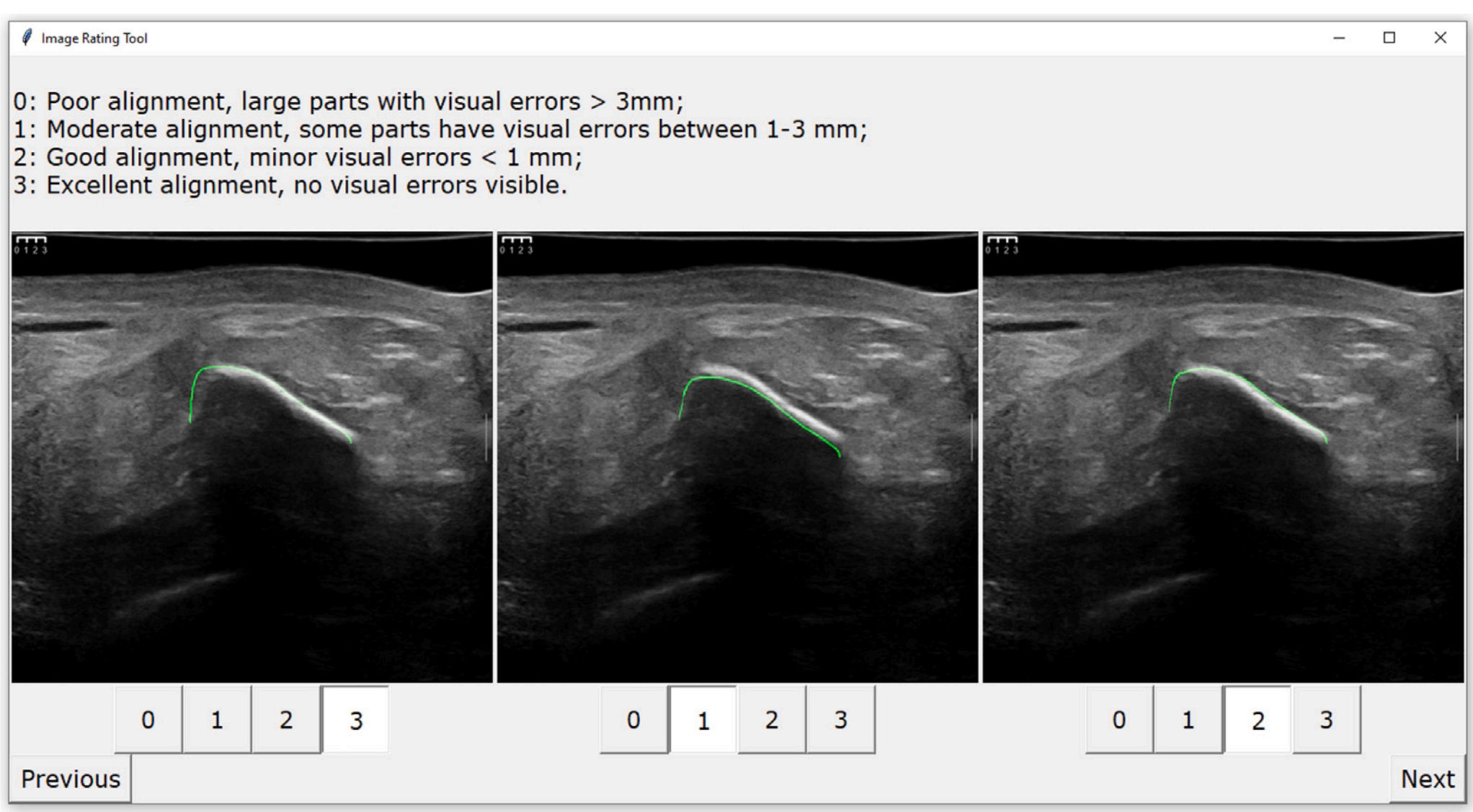

**Fig. 5.** A screenshot of the bone alignment rating software.

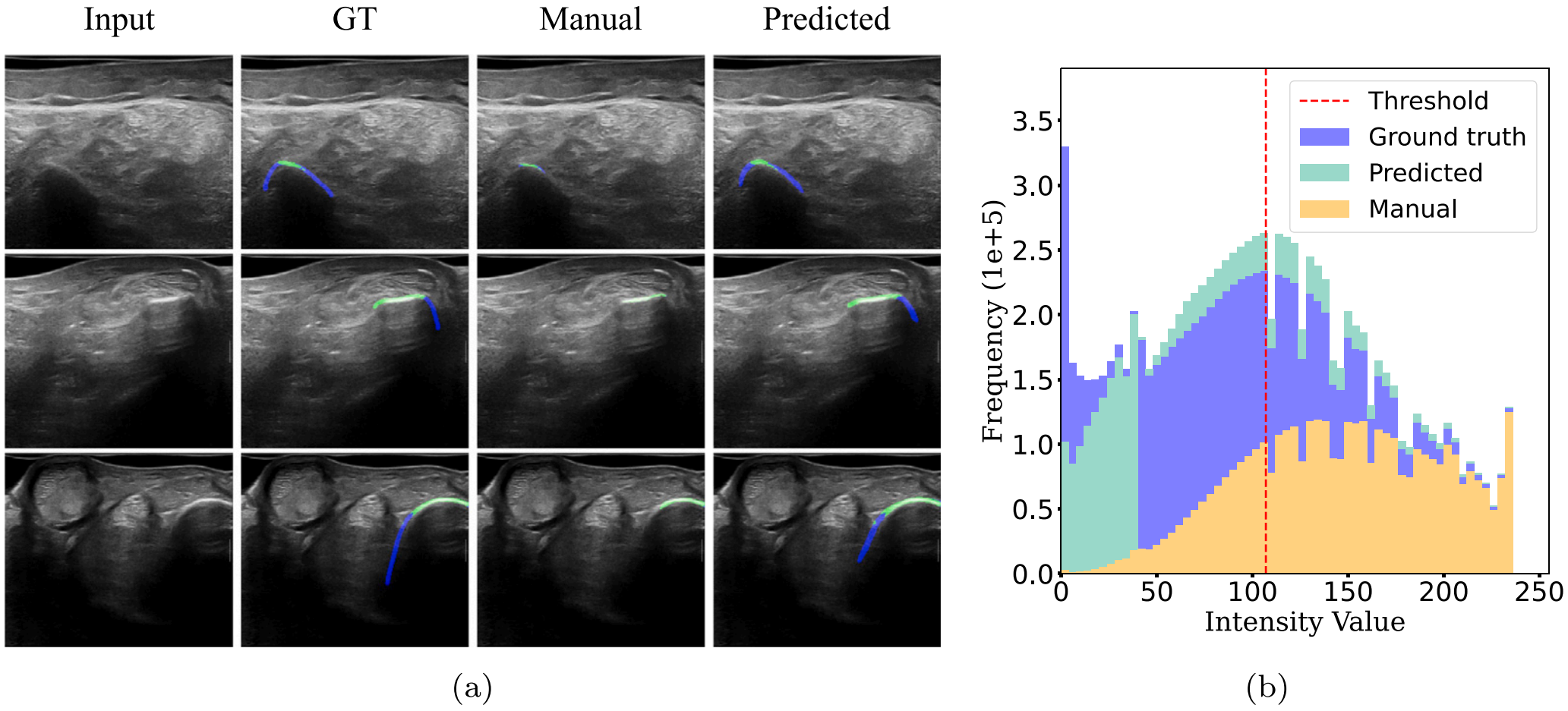

**Fig. 6.** (a) Qualitative comparison between manual labels and predicted ones. The first column shows the input images, the second column shows the CT-derived ground truth labels, the third column shows the manual labels, and the last column shows the bone labels predicted by the model trained on our labels. The bone labels are divided into high-intensity regions (green) and low-intensity regions (blue). (b) Distribution of pixel intensities over labels for manually labeled images. The red line represents the threshold determined by the Otsu method. While manual labels are mostly located in high-intensity regions, the proposed CT-derived labels allows for predictions in both high-intensity and low-intensity regions.

The accuracy, completeness, and F1 score with a distance threshold $\sigma$ are defined using Iverson brackets $[.]$, where $[Y]$ equals 1 if condition $Y$ is true, and 0 otherwise:

$$Acc(\sigma) = \frac{1}{|\mathcal{P}|} \sum_{\hat{\mathbf{x}} \in \mathcal{P}} [d_{\hat{\mathbf{x}} \to \mathcal{G}} < \sigma], \tag{4}$$

$$Com(\sigma) = \frac{1}{|\mathcal{G}|} \sum_{\mathbf{x} \in \mathcal{G}} [d_{\mathbf{x} \to \mathcal{P}} < \sigma], \tag{5}$$

$$F(\sigma) = \frac{2Acc(\sigma) \cdot Com(\sigma)}{Acc(\sigma) + Com(\sigma)}. \tag{6}$$

To particularly assess the performance of our method in segmenting lower-intensity bone regions, we quantitatively divide the labels into high-intensity and low-intensity regions. The intensity values of all bone pixels in the collected ultrasound images are recorded, and the Otsu method is applied to determine an intensity threshold separating the bone pixels into two classes: high-intensity and low-intensity [35]. The threshold is chosen to minimize intraclass variance and maximize interclass variance. Exemplary results are shown in Fig. 6.

**Generalizability.** Our dataset consists of images collected from the lower extremities of human subjects. To qualitatively assess the generalizability of the trained model to new anatomies and specimens, we apply it to ultrasound images of the pelvis and femur collected from new ex-vivo specimens. Qualitative results of the trained model's generalizability on ex-vivo anatomies are shown in Fig. 8. The model trained UltraBones100k successfully segment the bone surface in ultrasound images acquired from new bones, covering both the high intensity regions and low intensity regions. In addition, to qualitatively assess the model's generalizability on in-vivo ultrasound data, we apply it to in-vivo ultrasound images of the fibula and tibia. The results in Fig. 9 demonstrate that the model can accurately segment the bone surface in in-vivo ultrasound images, despite being trained exclusively on ex-vivo data.

## 4. Results

**Dataset Variability.** With our methodology, we have collected a dataset of 100k annotated ultrasound images with bone labels from 14 human cadaveric lower legs from different donors. The donors cover different physical characteristics, with heights ranging from 155 to 185 cm, weights from 59 to 99 kg, and BMI values between 19 and 30. The age range spans from 41 to 76 years, and the dataset includes both male and female specimens. Furthermore, arthritis status is documented, with some specimens having a history of falls or related conditions.

[7] https://www.imfusion.com/.

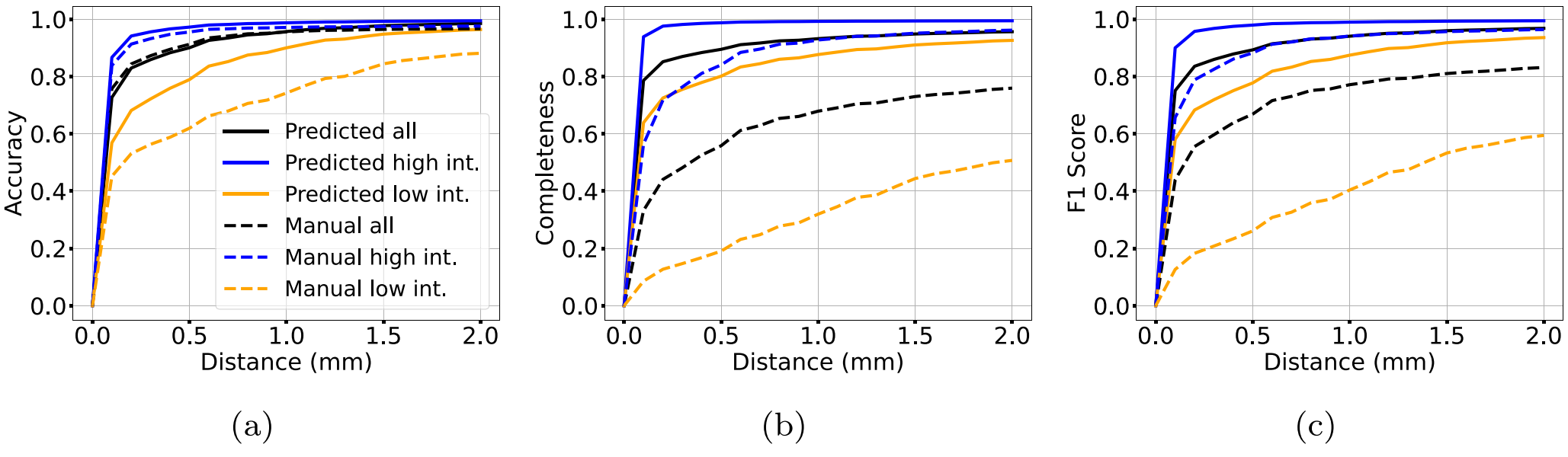

**Fig. 7.** Quantitative comparison between manual and predicted bone labels.

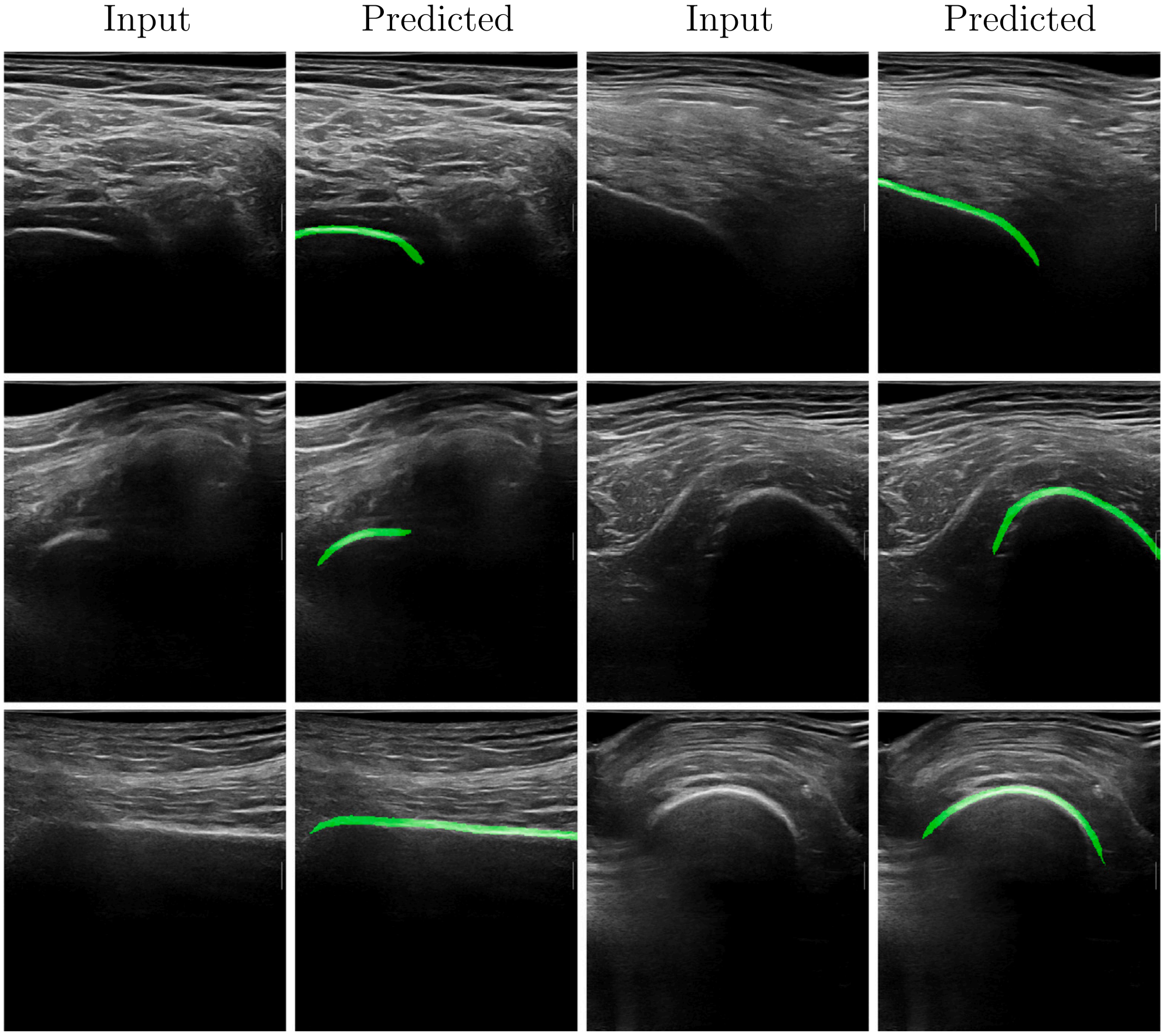

**Fig. 8.** Examples of ultrasound images depicting the pelvis and femur, along with the corresponding predicted bone labels.

**Table 2**
Results of the rating experiment. Each cell reports the mean value along with the standard deviation.

| Rater ID | Initial labeling | Global CT-US registration | Ours |
|---|---|---|---|
| Rater 1 | $1.45 \pm 0.66$ | $0.87 \pm 0.71$ | $2.39 \pm 0.76$ |
| Rater 2 | $1.83 \pm 0.67$ | $1.15 \pm 0.81$ | $2.56 \pm 0.62$ |

**Clinical Evaluation.** The results of the bone annotation rating experiment are presented in Table 2. For rater 1, the mean ratings are 1.45 for the initial alignment, 0.87 for CT-US registration, and 2.39 for our optimized alignment. For rater 2, the corresponding mean ratings are 1.83, 1.15, and 2.56. Wilcoxon signed-rank tests with Bonferroni correction indicate a significant improvement in bone labeling quality following our optimization pipeline ($p < 0.001$) for both raters.

**Automatic vs. Manual Labeling.** Qualitative results between the manual bone labels and the model-predicted labels are shown in Fig. 6. Quantitative results, including accuracy, completeness, and F1 score, are presented in Fig. 7. On average, at a distance threshold of 0.5 mm, the trained model achieves an accuracy of 0.900, completeness of 0.895, and an F1 score of 0.893 across all bone regions. Particularly, for low-intensity regions, the accuracy of the trained model (0.789) is 27.4% higher than that of manual labeling (0.619), the completeness (0.801) is 320% higher compared to manual labeling (0.190), and the F1 score (0.778) is 197% higher than the manual labeling (0.261).

## 5. Discussion

The results highlight our methodology as a promising solution for collecting large-scale ultrasound datasets with automatically generated bone labels. Clinical evaluation results confirm the quality of these labels. Furthermore, the femur and hip experiment results demonstrate the effectiveness of the collected dataset in training deep learning models with strong generalizability.

**Comparison with SOTA.** Previous work [12] aimed to align a tracked CT bone model with tracked ultrasound images. However, they reported that the mismatch of 1–2 mm was too large to allow for accurate automatic bone labeling in ultrasound images, and manual labeling was adopted in the end. In contrast, the results show that our optimization pipeline enhances CT-ultrasound alignment, enabling automatic accurate bone label generation. In addition, the applied global CT-US registration computes a single transformation matrix to align the original CT volume with the reconstructed ultrasound volume based on features such as intensity [23]. In the original CT volume, we observed local regions with high feature values, which can cause the optimization process to converge to local minima. As a result, the labeling quality becomes inconsistent: within the same ultrasound volume, some frames have high-quality labeling, while others perform poorly. Instead, our approach computes a transformation matrix for each individual ultrasound frame. According to the results of the clinical evaluation, our method achieves significantly improved bone annotations. This improvement is statistically validated by the Wilcoxon signed-rank test with Bonferroni correction ($p < 0.001$). However, our approach is significantly more computationally expensive and relies heavily on accurate local initialization from the initial tracking data.

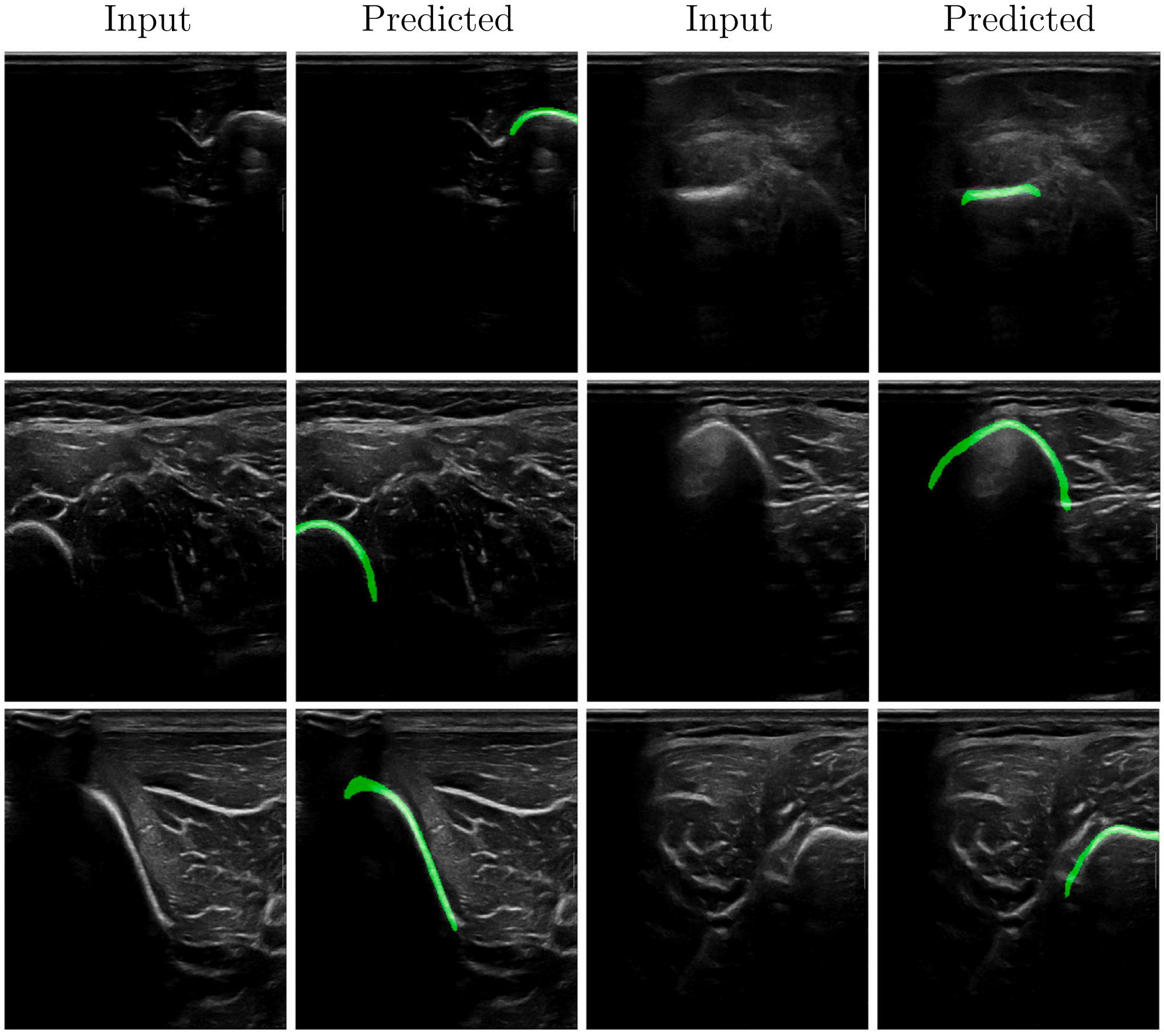

**Fig. 9.** Examples of ultrasound images depicting the in-vivo fibula and tibia, along with the corresponding predicted bone labels.

**Automatic vs. Manual Labeling.** Qualitative results in Fig. 6 indicate that the model trained on UltraBones100k outperforms the surgeon in segmenting bones in ultrasound images. Manual labels mostly include only high-intensity bone regions. In contrast, predicted bone labels include both high-intensity and low-intensity regions. Quantitative results in Fig. 7 demonstrate that the trained model consistently outperforms surgeon's manual labeling, regardless of the distance threshold used. Note that, compared to the trained model, manual labeling achieves comparable performance in terms of accuracy within the high-intensity regions. This result aligns well with our hypothesis that surgeons can accurately label bone surfaces in these regions. These qualitative and quantitative results in combination with the intensity distribution of labels illustrated in Fig. 6(b) further confirm that the expert surgeon faces difficulty in correctly labeling bone contours in low-intensity regions. These findings underscore the superiority of our approach over the manual labeling.

**Challenges in Dataset Availability.** Unlike natural images, manually labeling bones in ultrasound images requires specialized expertise in ultrasound imaging and bone anatomy, making it financially and temporally expensive. This constraint is a key factor limiting the size of training datasets in existing studies [22], which typically range from only dozens to a few thousand samples. The small dataset size restricts the performance and generalizability of trained models. Moreover, most existing datasets are not publicly available. As highlighted in prior reviews [21,22], the lack of a large-scale, publicly accessible dataset remains a major challenge in the ultrasound vision community. Our dataset collection methodology represents a solution to mitigate the dataset challenge. Furthermore, we make the proposed method, as well as the generated dataset publicly available. UltraBones100k is the largest known dataset with high quality bone labels, aiming to help advance future developments in intraoperative ultrasound for computer assisted surgery.

**Generalizability.** Although our dataset is the largest among existing works and includes specimens with diverse body conditions, its size remains relatively small compared to datasets in other domains. However, the fundamental principles of ultrasound imaging remain consistent across individuals, and the bone-soft tissue structure exhibits relatively uniform characteristics across different bodies. As a result, the appearance of bones in ultrasound images remains consistent. In the human body, bones are surrounded by soft tissues, and due to their high acoustic impedance, they consistently appear as bright regions in ultrasound images, regardless of anatomical differences. Additionally, since 2D ultrasound images capture only partial bone segments, they are often unaffected by variations in overall anatomical shape. The presence of shared structural features across bones—such as those in the arms and legs—further contributes to the model's generalizability. Qualitative results in Fig. 8 indicate the model's generalizability on ultrasound images of the pelvis and femur, even though it was trained on data collected from the lower extremities.

**Limitations.** Although the experiment results demonstrate the generalizability of the model trained on UltraBones100k, its performance may decline in anatomies with significantly different bone-soft tissue structures and shapes, such as the spine and skull. In addition, UltraBones100k consists of data only acquired from ex-vivo specimens. Compared to in-vivo specimens, the properties of biological tissues, such as elasticity and acoustic characteristics, change in cadaveric specimens, which can result in difference in ultrasound imaging. The distinction between ex-vivo and in-vivo ultrasound data is not a concern in scenarios that exclusively involve ex-vivo specimens, such as hands-on surgical skill training using cadaveric specimens [36]. However, for in-vivo applications, the difference can lead to a declined model performance. Although the results in Fig. 9 show that a model trained exclusively on ex-vivo data can accurately segment bone surfaces in in-vivo ultrasound images, additional in-vivo data collection may be necessary for fine-tuning the model [37,38], depending on the required level of accuracy. Conducting data collection studies with a limited number of patients is a common and justifiable approach in such scenarios. Furthermore, the publicly available model pretrained on UltraBones100k can be fine-tuned with newly collected data to enhance its performance for these specific cases. In addition, the segmentation performance can potentially be improved with network design and architecture optimization [39–41], particularly by considering the imaging characteristics of ultrasound. Although manual labeling is costly, obtaining annotations from multiple experts is valuable for comparison with automatic labeling and for evaluating both intra-rater and inter-rater variability.

## 6. Conclusion

Our methodology presents a promising solution to create large-scale and highly accurate training data for bone surface extraction in ultrasound images without relying on expert labeling. Serving as a benchmark, the segmentation model trained on UltraBones100k outperformed an expert surgeon with significantly more complete bone labels. The methodology and dataset show promise in advancing research and clinical translation of ultrasound imaging in computer-assisted interventions, such as surgical planning and intraoperative guidance, where 2D bone segmentation, 3D bone surface reconstruction, and multi-modality bone registration play critical roles.

## CRediT authorship contribution statement

**Luohong Wu:** Writing – review & editing, Writing – original draft, Visualization, Validation, Software, Methodology, Formal analysis, Data curation, Conceptualization. **Nicola A. Cavalcanti:** Writing – review & editing, Resources, Methodology, Investigation, Data curation, Conceptualization. **Matthias Seibold:** Writing – review & editing, Writing – original draft, Supervision, Project administration, Methodology, Investigation, Conceptualization. **Giuseppe Loggia:** Validation, Resources, Methodology, Conceptualization. **Lisa Reissner:** Validation, Resources, Methodology, Investigation. **Jonas Hein:** Writing – review & editing, Methodology. **Silvan Beeler:** Resources, Project administration. **Arnd Viehöfer:** Resources, Project administration, Methodology, Conceptualization. **Stephan Wirth:** Resources, Project administration. **Lilian Calvet:** Writing – review & editing, Writing – original draft, Supervision, Methodology. **Philipp Fürnstahl:** Writing – review & editing, Supervision, Resources, Project administration, Methodology, Funding acquisition, Conceptualization.

## Ethics statement

Ethics approval has been granted for our study by the Zurich cantonal ethics committee under BASEC Nr. 2023-01681. Additionally, all information regarding the human cadaveric specimens involved is anonymous.

## Declaration of Generative AI and AI-assisted technologies in the writing process

During the preparation of this work the authors used ChatGPT in order to improve the readability and language of the manuscript. After using this tool, the authors reviewed and edited the content as needed and take full responsibility for the content of the published article.

## Declaration of competing interest

All authors declare that they have no conflict of interest.

## Acknowledgments

This research has been funded by the Innosuisse Flagship project PROFICIENCY No. PFFS-21-19. This work has also been supported by the OR-X, a Swiss national research infrastructure for translational surgery, and associated funding by the University of Zurich and University Hospital Balgrist. We also appreciate Prof. Dr. med. Andreas Schweizer for his support and help.